\documentclass[12pt]{article}
\usepackage{amsmath}
\usepackage{amssymb}
\usepackage{graphicx}
\newcommand{\bb}{\begin{thebibliography}{99}}
\newcommand{\eb}{\end{thebibliography}}
\newcommand{\cL}{{\cal{L}}}

\title{ SCHWINGER-DYSON TYPE EQUATIONS FOR SOME QFT MODELS }
\author{ B.A.FAYZULLAEV,   E.QAYUMOV\\
 Department of  Theoretical Physics,\\
 National University of Uzbekistan,\\
 Tashkent 100095, Uzbekistan}

\begin{document}

\begin{titlepage}

 \maketitle

\begin{abstract}
The Schwinger-Dyson equations connecting free and full Green
functions and vertex parts widely were used in QED for finding full
Green functions under different conditions. Undoubtedly, the same
approach should leads to derivation of many useful information about
other models of QFT. In this work we present some technique based on
variational equations for effective action to derive many different
Schwinger-Dyson type equations in  QFT models such as nonlinear
sigma model and scalar field theory.

\end{abstract}
\vspace{.5cm} Key words: QED, Schwinger-Dyson equations, non-linear
sigma model, effective action, vacuum expectations, $n$-point
connected Green functions. \thispagestyle{empty}
\end{titlepage}
%\large

\section{Introduction}
Dyson \cite{ds}  and Schwinger \cite{scwh} had derived a system of equations which presents some integro-differential  relations between free and full Green functions
including vertex parts. On the base of these equations many results were obtained for QED full propagators under different conditions.
But this system of equations is not closed because for vertex parts it is impossible to find a finite system of equations.
This is due to that the basic equations are functional equations, and many functional relations may be obtained for different connected Green functions.
But derivation of these relations mainly is based on functional integration method and this approach connected with rather complicated consideration
\cite{fsch} - \cite{stenil}.  In this paper we present a method of derivation of the Schwinger-Dyson type equations based on simple differentiation
of equation for effective action. This approach follows to \cite{fm}, \cite{fb}  and \cite{fbconf}. We will see that following  this approach we can
derive many relations connecting different $n$-point Green functions for any QFT model.

\section{Some general relations}

Let's
to introduce the generator of all Green functions:
$$
Z[J]=\exp\{iW[J]\}=\int {\cal{D}}\varphi \exp\{i(S+J_i\varphi_i)\}.
$$
Here  the $\varphi_i$ is any field, $J_i$ is its external source,
$W[J]$ - is the generator of all connected Green's functions.
Hereafter  we will use the so-called condensed notations, for
example, $J_i\varphi_i$ means $J_i\varphi_i=\int d^4x
J(x)\varphi(x).$   Introducing so-called classical fields
\begin{equation}\label{clasfunc}
    \varphi_i=\frac{\delta W[J]}{\delta J_i}
\end{equation}
and performing following functional Legendre transformation
\begin{equation}\label{gamwjfi}
    \Gamma[\varphi]=W[J]-J_i\varphi_i
\end{equation}
we obtain the effective action $\Gamma[\varphi].$ According to
DeWitt \cite{dw}, Ch.22, the classical $S$ and quantum $\Gamma$
actions are connected as follows
\begin{equation}\label{dwrel}
    \frac{\delta\Gamma}{\delta\varphi_i}=\Lambda\frac{\delta S}{\delta\varphi_i},
\end{equation}
where the operator $\Lambda$ is constructed from connected Green
functions and functional derivatives over $\varphi_i$:
\begin{equation}\label{lambdaoper}
    \Lambda=:\exp\left\{\frac{i}{\hbar}\sum\limits_{n=2}^{\infty}\frac{(-i\hbar)^n}{n!}G_{i_1i_2\cdots
i_n}\frac{\delta^n}{\delta\varphi_{i_1}\delta\varphi_{i_2}\cdots\delta\varphi_{i_n}}\right\}:,
\end{equation}
where
$$
G_{ij}=\frac{\delta\varphi_i}{\delta J_j}=\frac{\delta^2W}{\delta
J_j \delta J_i},\qquad G_{i_1i_2\cdots i_n}=\frac{\delta^n W}{\delta
J_{i_1}\delta J_{i_2}\cdots \delta J_{i_n}}
$$
are connected (two- and $n$-point) Green functions. Commas in $\Lambda$ mean that derivatives act on r.h.s. expression only, not on $G'$s.
The
Eq.(\ref{dwrel}) connects $\Gamma$ and $n$-point Green functions
$G_{i_1i_2\cdots i_n}$, both of these are unknown quantity, so we
need in an additional relation for them. For this purpose we will
use following relation connecting the effective action $\Gamma$ and
the sources $J_i$ - so called quantum equations of motion (see
\cite{dw}):
\begin{equation}\label{delgamjot}
    \frac{\delta\Gamma}{\delta\varphi_i}=-J_i.
\end{equation}
But we can approach to the  Eq.(\ref{delgamjot}) from another point of view -  if rewrite Eq.(\ref{dwrel}) as follows
\begin{equation}\label{eqforw}
    -J_i=\Lambda  \frac{\delta S}{\delta\varphi_i},
\end{equation}
with the $\Lambda$ operator as in Eq.(\ref{lambdaoper}) then equations for $W[J]$ will be obtained.
This formula will be the main formula for us. We may rewrite it as follows:
$$
-J_i=\frac{\delta\Gamma}{\delta\varphi^i}=:\exp\{\frac{i}{\hbar}\sum_{n=2}^{\infty}\frac{(-i\hbar)^n}{n!}G^{i_1i_2\dots i_n}\frac{\delta^n}{\delta\varphi^i_1\delta\varphi^i_2\dots\delta\varphi^i_n}\}:\frac{\delta S}{\delta\varphi^i},
$$

\begin{figure}[h]
    \includegraphics[scale=0.45]{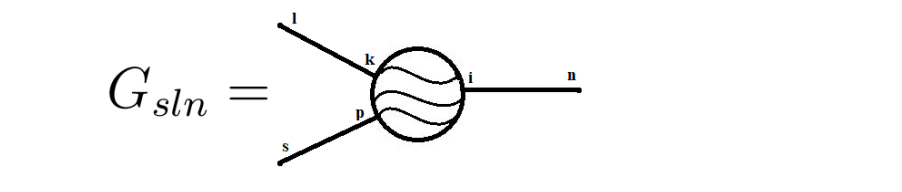}
    \caption{Three-point Green function}\label{pic1}
\end{figure}

Let's to connect three-point Green function and vertex part. For this we should to differentiate (\ref{delgamjot}) with respect $J_j:$
$$
\frac{\delta}{\delta J_j}\frac{\delta \Gamma}{\delta\varphi_i}=-\delta_{ij}.
$$
Due to
$$
\frac{\delta\varphi_k}{\delta J_l}=\frac{\delta}{\delta J_l}\frac{\delta W}{\delta J_k}
=\frac{\delta^2W}{\delta J_l\delta J_k}=G_{lk}
$$
we get
$$
\frac{\delta^2W}{\delta J_l\delta J_k}\frac{\delta^2\Gamma}{\delta\varphi_k\delta\varphi_i}=-\delta_{il}
$$
Differentiating once more over  $\frac{\delta}{\delta J_s}$ we get:
\begin{equation}\label{d3wjjj}
    \frac{\delta^3W}{\delta J_s\delta J_l\delta J_n}=\frac{\delta^2W}{\delta J_l\delta J_k}\frac{\delta^2W}{\delta J_p\delta J_s}\frac{\delta^2W}{\delta J_i\delta J_n}\frac{\delta^3\Gamma}{\delta\varphi_p\delta\varphi_k\delta\varphi_i}
\end{equation}
So we find a relation which connects three-point connected Green function with three-point vertex part:
\begin{equation}\label{glsn}
    G_{lsn}=G_{lk}G_{sp}G_{ni}\Gamma_{kpi},\qquad \Gamma_{kpi}=\frac{\delta^3 \Gamma}{\delta\varphi_k\delta\varphi_p\delta\varphi_i}.
\end{equation}
This is depicted in the Fig.\ref{pic1}.
We may continue and once more differentiate Eq.(\ref{d3wjjj}), what gives us the following relation between four-  and three- and two-point Green functions:
\begin{multline*}
\frac{\delta^4W}{\delta J_m\delta J_s\delta J_l\delta J_n}=\frac{\delta^3W}{\delta J_m\delta J_l\delta J_k}\frac{\delta^2W}{\delta J_p\delta J_s}\frac{\delta^2W}{\delta J_i\delta J_n}\frac{\delta^3\Gamma}{\delta\varphi_p\delta\varphi_k\delta\varphi_i}+\\+\frac{\delta^2W}{\delta J_l\delta J_k}\frac{\delta^3W}{\delta J_m\delta J_p\delta J_s}\frac{\delta^2W}{\delta J_i\delta J_n}\frac{\delta^3\Gamma}{\delta\varphi_p\delta\varphi_k\delta\varphi_i}
+\frac{\delta^2W}{\delta J_l\delta J_k}\frac{\delta^2W}{\delta J_p\delta J_s}\frac{\delta^2W}{\delta J_m\delta J_i\delta J_n}\frac{\delta^3\Gamma}{\delta\varphi_p\delta\varphi_k\delta\varphi_i}+\\+\frac{\delta^2W}{\delta J_l\delta J_k}\frac{\delta^2W}{\delta J_p\delta J_s}\frac{\delta^2W}{\delta J_i\delta J_n}\frac{\delta^2W}{\delta J_m\delta J_d}\frac{\delta^4\Gamma}{\delta\varphi_d\delta\varphi_p\delta\varphi_k\delta\varphi_i}
\end{multline*}
or,
\begin{equation}\label{key4}\begin{array}{l}
\displaystyle{G_{msln}=G_{mlk}G_{ps}G_{in}\Gamma_{pki}+G_{lk}G_{mps}G_{in}\Gamma_{pki}+}\\ \\ \displaystyle{G_{lk}G_{ps}G_{min}\Gamma_{pki}+G_{lk}G_{ps}G_{in}G_{md}\Gamma_{dpki}}.
\end{array}
\end{equation}
This expression ay be depicted as in the Fig:
    \begin{figure}[h]
        \includegraphics[scale=0.45]{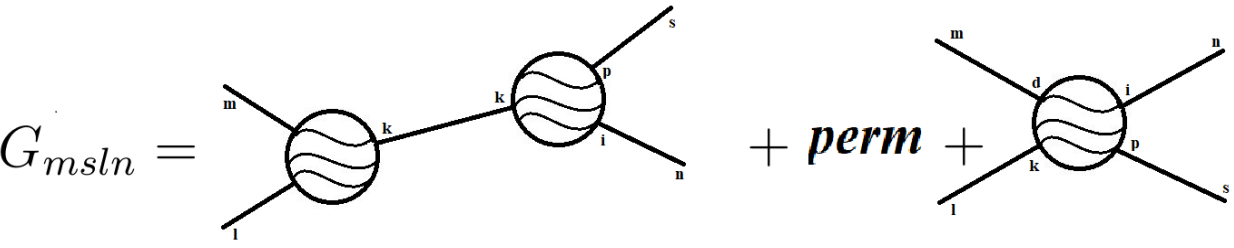}
    \caption{Four-, three- and two-point functions.}\label{pic2}
        \end{figure}

\section{Scalar $\lambda \varphi^4$ theory }

\begin{figure}[h]
        \includegraphics[scale=0.45]{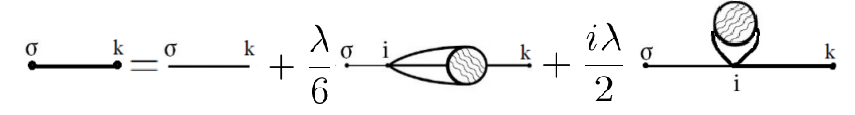}
\caption{Graphical representation of the Schwinger-Dyson equation for $\lambda\varphi^4/4$ theory}\label{pic3}
    \end{figure}

For example, for scalar $\lambda\varphi^4$ theory classical action looks like:
$$
S=-\frac12 \varphi_i(\partial^2+m^2)_{ij}\varphi_j-\lambda \varphi^4=-\frac12 \varphi_i K^{-1}_{ij}\varphi_j-\lambda \varphi_i^4.
$$
Applying to this classical action Eq.(\ref{eqforw})  with $\Lambda$  from (\ref{lambdaoper}) one can obtains equation for $W:$
\begin{equation}\label{lamphi4}
\frac{\lambda}{6}\hbar^2\frac{\delta^3W}{\delta J_i^3}+\frac{i\lambda\hbar}{2}\frac{\delta^2W}{\delta J_i^2}\frac{\delta W}{\delta J_i}-\frac{\lambda}{6}\left(\frac{\delta W}{\delta J_i}\right)^3-K^{-1}_{ij}\frac{\delta W}{\delta J_j}+J_i=0.
\end{equation}
where
$$
K^{-1}_{ij}=-i\left(\partial^2+m^2\right)^{-1}\delta_{ij}
$$
is free Green function. Differentiating Eq.(\ref{lamphi4}) over
$\dfrac{\delta}{\delta J_k}$ we get:
        \begin{multline*}
    \frac{\hbar^2\lambda}{6}\frac{\delta^4W}{\delta J_k\delta J_i\delta J_i\delta J_i}-\frac{i\hbar \lambda}{2}\left(\frac{\delta^3W}{\delta J_k\delta J_i \delta J_i}\varphi_i+\frac{\delta^2W}{\delta J_i \delta J_i}\frac{\delta^2W}{\delta J_k \delta J_i}\right)-\\-\frac{\lambda}{2}\varphi^2_i\frac{\delta^2W}{\delta J_k \delta J_i}-K^{-1}_{ij}\frac{\delta^2W}{\delta J_k \delta J_j}=-\delta_{ki}.
\end{multline*}
Multiplying this equation by
     $K_{\sigma i} $ we may obtain:
\begin{multline*}
    K_{\sigma i}\frac{\hbar^2\lambda}{6}\frac{\delta^4W}{\delta J_k\delta J_i\delta J_i\delta J_i}-K_{\sigma i}\frac{i\hbar \lambda}{2}\left(\frac{\delta^3W}{\delta J_k\delta J_i \delta J_i}\varphi_i+\frac{\delta^2W}{\delta J_i \delta J_i}\frac{\delta^2W}{\delta J_k \delta J_i}\right)-\\
    -K_{\sigma i}\frac{\lambda}{2}\varphi^2_i\frac{\delta^2W}{\delta J_k \delta J_i}-K_{\sigma i}K^{-1}_{ij}\frac{\delta^2W}{\delta J_k \delta J_j}=-K_{\sigma i}\delta_{ki}
    \end{multline*}
Putting
     $(J=0)$ we find equation for full Green function - the Schwinger-Dyson equation:
    \begin{equation}\label{yechim}
    G^{\sigma k}=K_{\sigma k}+\frac{\hbar^2\lambda}{6}K_{\sigma i}G^{kiii}+\frac{i\hbar \lambda}{2}K_{\sigma i}(G^{kii}\varphi_i+G^{ii}G_{ki})-\frac{\lambda}{2}\varphi^2_iG^{ki}K_{\sigma i}
    \end{equation}
This equation is presented in the Fig.\ref{pic3}.

\section{The case of QED}
\begin{figure}[h]
        \includegraphics[scale=0.45]{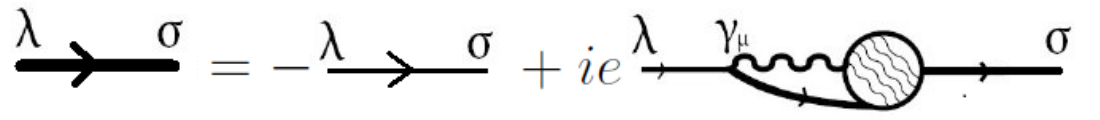}
        \caption{
Eq.(\ref{elektro2})-standard Schwinger-Dyson equation for full electron Green function}\label{pic8}
    \end{figure}

In \cite{fm} the following set of equations for QED effective action was derived:
\begin{equation}\label{elektro1} -J_{\mu}=e\bar{\Psi}\gamma^\mu\Psi +D^{-1\mu\nu}A_{\nu}+ie\hbar \mathrm{Tr}\left(\gamma^\mu\frac{\delta^2W}{\delta\eta\delta\bar{\eta}}\right).
\end{equation}
\begin{equation}\label{elektro2}
-\eta_{\alpha}=((-\hat\partial-m+e\hat A)\Psi)^{\alpha}-ie\hbar(\gamma_\mu)^{\alpha\beta}\frac{\delta^2W}{\delta J_\mu\delta\bar{\eta}^\beta}.
\end{equation}
\begin{equation}\label{elektro3}
\bar{\eta}_\alpha=\left(\bar{\Psi}\left(i\overleftarrow{\hat{\partial}}-e\hat{A}+m \right)\right)^\alpha-ie\hbar\left(\gamma_\mu\right)^{\beta\alpha}\frac{\delta^2W}{\delta J_\mu\delta\eta^\beta}.
\end{equation}
Here $J_mu, \, \eta_i$ and $\bar{\eta}_i$ are classical sources of the fields $A_\mu, \bar{\psi}_i$ and $\psi_i$, consequently, and
$$
D_{\lambda\mu}=\frac{1}{\partial^2}\left(g_{\lambda\mu}-(1-\alpha)\frac{\partial_\lambda\partial_\mu}{\partial^2}
\right)
$$
is free propagator of the electromagnetic field.
Differentiating Eq.(\ref{elektro1}) over  $\dfrac{\delta}{\delta J_\sigma}$ and multiplying the result by $D_{\lambda\mu}$  gives us:
\begin{multline*}
-D_{\lambda\mu}\delta_{\sigma\mu}=eD^{\lambda\mu}\frac{\delta\bar\Psi}{\delta J_{\sigma}}\gamma^\mu\Psi+eD_{\lambda\mu}\bar\Psi\gamma^\mu\frac{\delta\Psi}{\delta J_\sigma}+D_{\lambda\mu}D^{-1\mu\nu}\frac{\delta^2W}{\delta J_\sigma\delta J_\nu}+\\+ie\hbar D_{\lambda\mu} Tr\left(\gamma^\mu\frac{\delta^3W}{\delta J_\sigma\delta\eta\delta\bar{\eta}}\right).
\end{multline*}
In the source-free case  ( $J_\mu=0 ; \eta=0 ; \bar{\eta}=0$ ) we get
$$
G_{\lambda\sigma}=-D_{\lambda\sigma}-ie\hbar D_{\lambda\mu} Tr\left(\gamma^\mu\frac{\delta^3W}{\delta J_\sigma\delta\eta\delta\bar{\eta}}\right).
$$
Here according to Eq.(\ref{glsn}) we should write down:
$$
\frac{\delta^3W}{\delta J_\sigma\delta\eta_\alpha\delta\bar{\eta}_\beta}=G_{\sigma\nu}G^{\alpha\rho}G^{\beta\\tau}\frac{\delta^3 \Gamma}{\delta A_\nu\delta\psi^\rho\delta\bar{\psi}^\tau}.
$$
Here $G_{\sigma\nu}$ is full Green function of the photon, $G^{\alpha\beta}$ - are full Green functions of the electron. The last term is three-point vertex part.
This expression is standard Schwinger-Dyson equation for QED. If the vertex part may be presented in the form
$$
G^{\nu}_{\rho\tau}=\frac{\delta^3 \Gamma}{\delta A_\nu\delta\psi^\rho\delta\bar{\psi}^\tau}
$$
then the Schwinger-Dyson equation may be presented as follows:
\begin{equation}\label{elektro/y}
G_{\lambda\sigma}=-D_{\lambda\sigma}-ie\hbar D_{\lambda\mu} Tr\left(\gamma^\mu G^{\sigma}\right).
\end{equation}
    \begin{figure}[h]
        \includegraphics[scale=0.45]{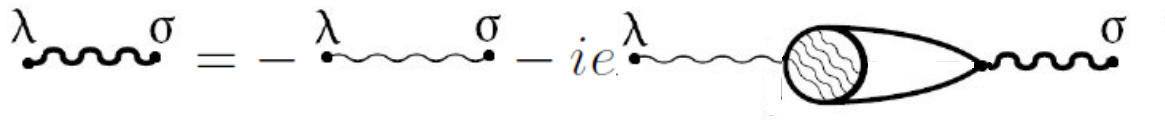}
        \caption{Schwinger-Dyson equation for QED}
    \end{figure}
In principle, this relation allows to calculate the full Green function of photon if we know all other terms.
Let's pass on to equation for full Green function of the electron.
\begin{equation}
-\eta_{\alpha}=((-\hat\partial-m+e\hat A)\Psi)^{\alpha}-ie\hbar(\gamma_\mu)^{\alpha\beta}\frac{\delta^2W}{\delta J_\mu\delta\bar{\eta}^\beta}
\end{equation}
Differentiating this equation by  $\dfrac{\delta}{\delta\eta_\sigma}$ we get:
$$
-\delta_{\alpha\sigma}=(i\hat{\partial}-m)^{\alpha\beta}\frac{\delta\Psi_\beta}{\delta\eta_\sigma}+e\frac{\delta\left((\hat{A})^{\alpha\beta}\Psi_\beta\right)}{\delta\eta_\sigma}
-ie\hbar(\gamma_\mu)^{\alpha\beta}\frac{\delta^3W}{\delta\eta_\sigma\delta J_\mu \delta\bar{\eta}^\beta}
$$
or, after somemanipulation
we have standard Schwinger-Dyson equation:
\begin{equation}\label{elekt/y2}
G^{\lambda\sigma}=-S_{\lambda\sigma}+ie\hbar S_{\lambda\alpha}(\gamma_\mu)^{\alpha\beta}G^\mu_{\beta\sigma}.
\end{equation}
This relation is depicted in Fig.\ref{pic8}.

\section{Nonlinear $\sigma$-model}

\begin{figure}[h]
\includegraphics[scale=0.45]{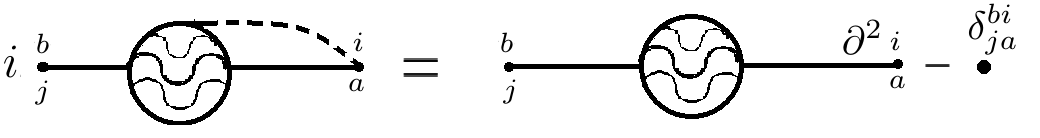}
\caption{Feynman diagram for Eq.\ref{Gbijai}}\label{pic5}
\end{figure}

Let's consider  following model
$$
\mathcal{L}=\frac12 (\partial_\mu\boldsymbol{\sigma})^2
$$
where
 a $N$-component scalar field $\{\sigma_a(x),\,\,a=1, 2, ..., N\}$
 is subject to the constraint
\begin{equation}\label{constr}
    \boldsymbol{\sigma}^2(x)=\sum\limits_{a=1}^N\sigma_a(x)\sigma_a(x)=\frac{N}{\gamma}.
\end{equation}
Although at first sight in the model there is no interaction but
solving the constraint Eq.(\ref{constr}) with respect to one of the
components we arrive at non-trivial interaction between remaining
components. The coefficient $\gamma$ turns out to be a coupling
constant. We can take into account this nontrivial structure of the
model by introduction of an auxiliary field - a Lagrange multiplier
- $\alpha(x)$ by the following way:
\begin{equation}\label{mod}
    \cL=\frac12 (\partial_\mu\boldsymbol{\sigma})^2-\frac12\alpha\left(\boldsymbol{\sigma}^2-\frac{N}{\gamma}\right),\qquad \boldsymbol{\sigma}=\{\sigma^a,\,a=1, 2, 3, \, ... ,\, N\},
\end{equation}
where  $\alpha(x)$ - is an auxiliary scalar field. In the condensed
notations we have for the action:
$$
S=-\frac12\sigma^a_i(\partial^2_i+\alpha_i)\sigma^a_i+\frac{N}{2\gamma}\alpha_i=-\frac12\sigma^a_iD_{ij}
\sigma^a_j +\frac{N}{2\gamma}\alpha_i,
$$
where $D_{ij}=\left(\partial^2_i+\alpha_i\right)\delta_{ij}$.
As it was shown in \cite{fb} the equations for effective action for this model has following form:
\begin{equation}
i\hbar\frac{\delta^2 W}{\delta j^i_a\delta\eta_i}=\partial^2\frac{\delta W}{\delta j^i_a}+\frac{\delta W }{\eta_i}\frac{\delta W}{\delta j^i_a}-j^i_a
\label{jia}
\end{equation}
\begin{equation}
i\hbar\frac{\delta^2 W}{\delta j^i_a\delta j^i_a}=\left(\frac{\delta W}{\delta j^i_a}\right)^2-\frac{N}{\gamma}-2\eta_i
\label{etai}
\end{equation}
Further we will work with these equations.
Let's to differentiate Eq.(\ref{jia})  over $j^b_j$, and Eq.(\ref{etai})-over $\eta_j$:
\begin{equation}
i\frac{\delta^3 W}{\delta j^b_j\delta j^i_a\delta\eta_i}=\partial^2\frac{\delta^2 W}{\delta j^b_j\delta j^i_a}+\frac{\delta^2W}{\delta j^b_j\delta\eta_i}\frac{\delta W}{\delta j^i_a}+\frac{\delta W}{\delta\eta_i}\frac{\delta^2W}{\delta j^b_j\delta j^i_a}-\delta^{bi}_{ja}
\label{2-hosila}
\end{equation}
\begin{equation}
i\frac{\delta^3W}{\delta\eta_j\delta j^i_a\delta j^i_a}=2\frac{\delta W}{\delta j^i_a}\frac{\delta^2W}{\delta\eta_j\delta j^i_a}-2\delta_{ji}\label{2-hosila2}
\end{equation}
Denoting
$$
G^{ib}_{aj}=\frac{\delta^2W}{\delta j^i_a\delta j^b_j},\qquad G^a_{ij}=\frac{\delta^2 W}{\delta j^a_i\delta
\eta_j}
$$
and differentiating Eq.(\ref{jia})over  $j^b_j$, and  Eq. (\ref{etai})- over  $\eta_j$ we get the following relations for full connected Green functions:
\begin{equation}\label{secgabij}\begin{array}{l}
iG^{bi*}_{jai}=\partial^2 G^{bi}_{ja}+G^{b*}_{ji}\sigma^i_a+\alpha_i G^{bi}_{ja}-\delta^{bi}_{ja};\\
iG^{*ii}_{jaa}=2\sigma^i_aG^{*i}_{ja}-2\delta_{ji}.
\end{array}\end{equation}
In the source-free case and supposing $\dfrac{\delta W}{\delta j^i_a}=0; \dfrac{\delta W}{\eta_i}=0
$  we obtain the first Schwinger-Dyson type equation:
\begin{equation}\label{Gbijai}
iG^{bi*}_{jai}=\partial^2 G^{bi}_{ja}-\delta^{ij}\delta_{ab}.
\end{equation}
This relation may be depicted as in the Fig.\ref{pic5}.
From the second of Eq.(\ref{secgabij}) we may obtain (in the source free case)
\begin{equation}\label{ddiijaa}
G^{*ii}_{jaa}=2i\delta_{ji}.
\end{equation}

    \begin{figure}[h]
        \includegraphics[scale=0.45]{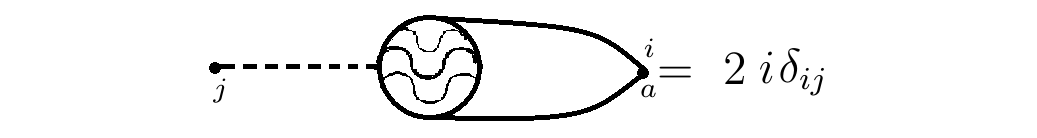}
\caption{Graphic representation for Eq.\ref{ddiijaa})}
    \end{figure}
\begin{figure}[h]
        \includegraphics[scale=0.45]{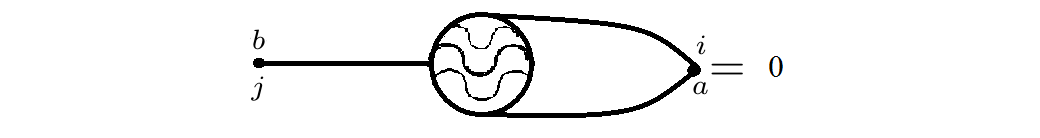}
\caption{Graphical representation of the Eq.(\ref{biijaa}).}\label{pic6}
    \end{figure}
\begin{figure}[h]
\includegraphics[scale=0.45]{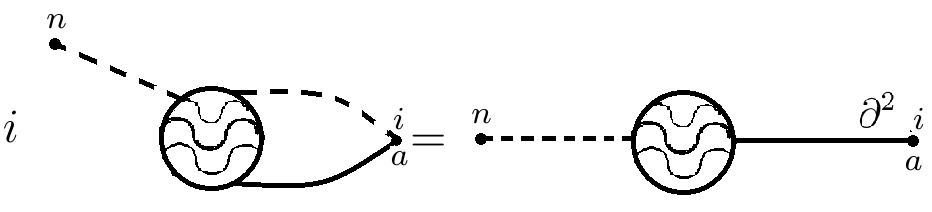}
\caption{Eq.(\ref{nai}) in diagrammatic form.}\label{pic7}
\end{figure}
\begin{figure}[h]
        \includegraphics[scale=0.45]{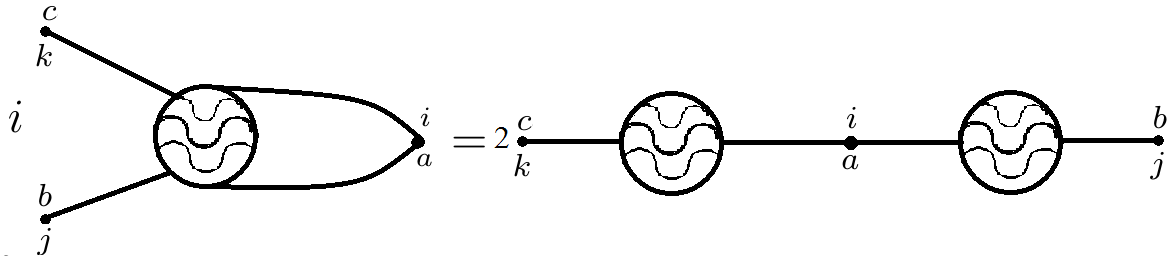}
        \caption{Eq.(\ref{cbiikj}) in diagrammatic form}\label{pic9}
    \end{figure}
\begin{figure}[h]
        \includegraphics[scale=0.5]{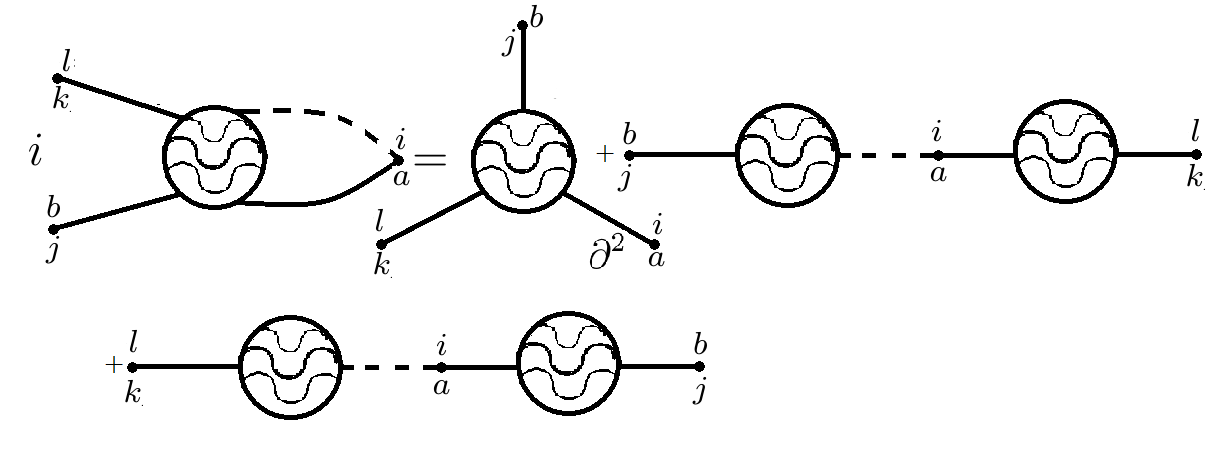}
\caption{Eq.(\ref{lbi*}) in graphic form}\label{pic10}
    \end{figure}
\begin{figure}
\includegraphics[scale=0.5]{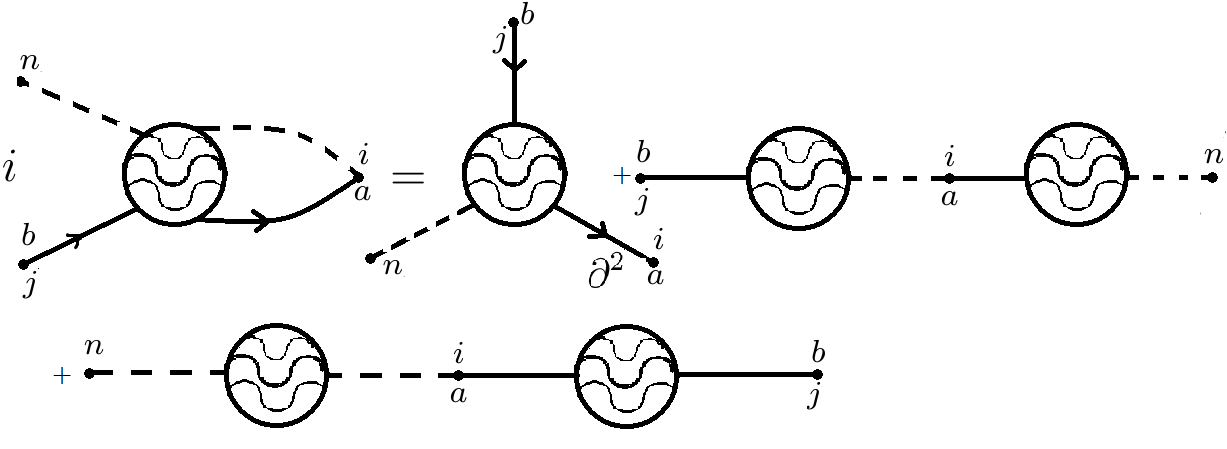}
\caption{Eq.(\ref{*bi*})}\label{pic11}
\end{figure}

    After differentiating of Eq.(\ref{etai}) over  $j^b_j$ we get
\begin{equation}\label{etaibj}
i\frac{\delta^3 W}{\delta j^b_j\delta j^i_a\delta j^i_a}=2\frac{\delta W}{\delta j^i_a}\frac{\delta^2 W}{\delta j^b_j\delta j^i_a},
\end{equation}
 what may be presented as follows:
$$
G^{bii}_{jaa}=2\sigma^i_aG^{bi}_{ja}.
$$
If we put sources equal to zero and suppose $\sigma_i^a=0$ in this case then
\begin{equation}\label{biijaa}
G^{bii}_{jaa}=0.
\end{equation}
This is presented in the Fig.\ref{pic6}.
Differentiating of the Eq.(\ref{jia}) over  $\eta_n$ gives us:
$$
i\frac{\delta^3W}{\delta\eta_n\delta j^i_a\delta\eta_i}=\partial^2\frac{\delta^2W}{\delta\eta_n\delta j^i_a}+\frac{\delta^2W}{\delta\eta_n\delta\eta_i}\frac{\delta W}{\delta j^i_a}+\frac{\delta W}{\delta\eta_i}\frac{\delta^2W}{\delta\eta_n\delta j^i_a}.
$$
Passing to Green functions we may present this as follows:
$$
iG^{*i*}_{nai}=\partial^2G^{*i}_{na}+G^{**}_{ni}\sigma^i_a+\alpha_iG^{*i}_{na}.
$$
Again turning to the source-free case we have:
\begin{equation}\label{nai}
iG^{*i*}_{nai}=\partial^2G^{*i}_{na}.
\end{equation}
what is presented in Fig.\ref{pic7}.
Let's to differentiate Eq.(\ref{etaibj}) over $j^c_k$:
$$
i\frac{\delta^4W}{\delta j^c_k\delta j^b_j\delta j^i_a\delta j^i_a}=2\frac{\delta^2W}{\delta j^c_k\delta j^i_a}\frac{\delta^2 W}{\delta j^b_j\delta j^i_a}+2\frac{\delta W}{\delta j^i_a}\frac{\delta^3 W}{\delta j^c_k\delta j^b_j\delta j^i_a},
$$
or, through Green functions in the presence of sources:
$$
iG^{cbii}_{kjaa}=2G^{ci}_{ka}G^{bi}_{ja}+2\sigma^i_aG^{cbi}_{kja}
$$
In the source-free case:
\begin{equation}\label{cbiikj}
iG^{cbii}_{kjaa}=2G^{ci}_{ka}G^{bi}_{ja},
\end{equation}
which is presented in the Fig.\ref{pic9}.
Taking derivative on  $ j^l_k$ from Eq. (\ref{2-hosila}) gives us
\begin{multline*}
i\frac{\delta^4W}{\delta j^l_k\delta j^b_j\delta j^i_a\delta\eta_i}=\partial^2\frac{\delta^3W}{\delta j^l_k\delta j^b_j\delta j^i_a}+\frac{\delta^3W}{\delta j^l_k\eta_i\delta j^b_j}\frac{\delta W}{\delta j^i_a}+\frac{\delta^2W}{\delta j^b_j\delta\eta_i}\frac{\delta^2W}{\delta j^l_k\delta j^i_a}+\\+\frac{\delta^2W}{\delta j^l_k\delta\eta_i}\frac{\delta^2W}{\delta j^b_j\delta j^i_a}+\frac{\delta W}{\delta\eta_i}\frac{\delta^3W}{\delta j^l_k\delta j^b_j\delta j^i_a},
\end{multline*}
or,
$$
iG^{lbi*}_{kjai}=\partial^2G^{lbi}_{kja}+G^{l*b}_{kij}\sigma^i_a+G^{b*}_{ji}G^{li}_{ka}+G^{l*}_{ki}G^{bi}_{ja}+\alpha_iG^{lbi}_{kja}.
$$

\begin{equation}\label{lbi*}
iG^{lbi*}_{kjai}=\partial^2G^{lbi}_{kja}+G^{b*}_{ji}G^{li}_{ka}+G^{l*}_{ki}G^{bi}_{ja}
\end{equation}
This is presented in Fig.\ref{pic10}.
 Now we will take derivative of Eq.(\ref{2-hosila}) over $\eta_n$, this gives us
\begin{equation}
\begin{array}{l}
\displaystyle{i\frac{\delta^4 W}{\delta\eta_n\delta j^b_j\delta j^i_a\delta\eta_i}=\partial^2 \frac{\delta^3 W}{\delta\eta_n\delta j^b_j\delta j^i_a}+\frac{\delta^3 W}{\delta\eta_n\delta j^b_j\delta\eta_i}\frac{\delta W}{\delta j^i_a}}+\\ \\ \displaystyle{+\frac{\delta^2 W}{\delta j^b_j\delta\eta_i}\frac{\delta^2 W}{\delta\eta_n\delta j^i_a}+\frac{\delta^2 W}{\delta\eta_n\delta\eta_i}\frac{\delta^2 W}{\delta j^b_j\delta j^i_a}+\frac{\delta W}{\delta\eta_i}\frac{\delta^3W}{\delta\eta_n\delta j^b_j\delta j^i_a}},
\end{array}
\label{etan}
\end{equation}
or,
\begin{equation}
iG^{*bi*}_{njai}=\partial^2 G^{*bi}_{nja}+G^{*b*}_{nji}\sigma^i_a+G^{b*}_{ji}G^{*i}_{na}+G^{**}_{ni}G^{bi}_{ja}+\alpha_i G^{*bi}_{nja}.
\end{equation}
In source-free case:
\begin{equation}\label{*bi*}
iG^{*bi*}_{njai}=\partial^2 G^{*bi}_{nja}+G^{b*}_{ji}G^{*i}_{na}+G^{**}_{ni}G^{bi}_{ja}.
\end{equation}
This relation is presented in Fig.\ref{pic11}.
Once more differentiating Eq.(\ref{2-hosila2}) over $\eta_k$
we get:
$$
i\frac{\delta^4W}{\delta\eta_k\delta\eta_j\delta j^i_a\delta j^i_a}=2\frac{\delta^2 W}{\delta\eta_k\delta j^i_a}\frac{\delta^2W}{\delta\eta_j\delta j^i_a}
+2\frac{\delta W}{\delta j^i_a}\frac{\delta^3W}{\delta\eta_k\delta\eta_j\delta j^i_a}.
$$
This means, that
$$
iG^{**ii}_{jkaa}=2G^{*i}_{ka}G^{*i}_{ja}+2\sigma^i_aG^{**i}_{kja}.
$$
In source-free case we have:
\begin{equation}\label{iijkaa}
iG^{**ii}_{jkaa}=2G^{*i}_{ka}G^{*i}_{ja}.
\end{equation}
In graphic form this is presented in Fig.\ref{pic12}:
\begin{figure}[h]
\includegraphics[scale=0.45]{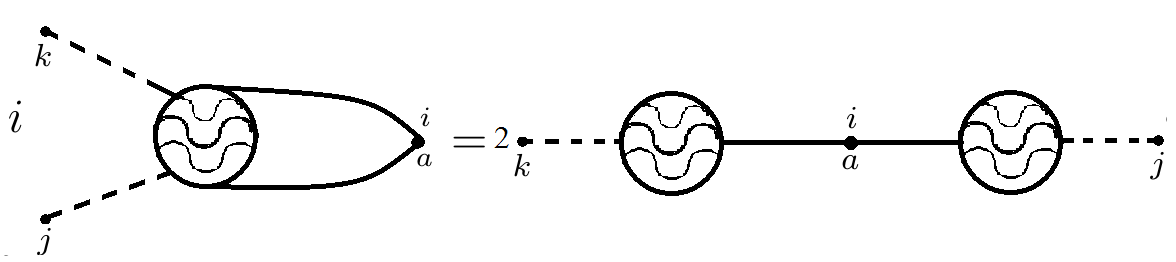}
\caption{Eq.(\ref{iijkaa})}\label{pic12}
\end{figure}
\section{Conclusion}

We have shown that the method of variational equations for effective action gives us a powerful tool for derivation of many relations between different
 Green functions
of any QFT models. For this derivation it is sufficient to differentiate the main equation for effective action in any QFT model
as many times as it is necessary.


\begin{thebibliography}{99}
\bibitem{ds} F.Dyson, Phys.Rev., v.75, 1736 (1949).
\bibitem{scwh}  J.Schwinger, Procl. Natl. Acad. Sci., v.37, p.452(1951).
\bibitem{dw} B. S. DeWitt. "Dynamical Theory of Groups and Fields," Gordon and Breach, New York, 1965.
\bibitem{fm} B.A.Fayzullaev and M.M.Musakhanov, Two-loop effective action for theories with fermions, Annals of Phys.(NY) {\bf 241} (1995)394.
\bibitem{fb} B.A.Fayzullaev, Effective action and vacuum expectations for nonlinear $\sigma$-model, arXiv:1510.07367.
\bibitem{fbconf} B.A.Fazullaev, Int.Journ.of Modern Physics, Conference Series, v.49 (2019)1960006.
\bibitem{fsch} Fischer C.S. Infrared properties of QCD from Dyson-Schwinger equations. \textit{J.Phys.} \underline{G32} (2006)253-291.
\bibitem{fischalk} C. S. Fischer and Alkofer \textit{Phys. Rev} \textbf{D67} (2003) 094020, hep-ph/0301094
\bibitem{mr} Maris P., Roberts C.D. Dyson-Schwinger equations: A Tool for hadronic physics. \textit{Int.J.Mod.Phys.} \underline{E12} (2003)297-365.
\bibitem{robschm} Roberts C.D., Schmidt S.M. Dyson-Schwinger equations: Density, temperature and continuum strong QCD. \textit{Prog.Part.Nucl.Phys.}, \underline{45} (2000) 1-103.
\bibitem{robwill} Roberts C.D., Williams A. Dyson-Schwinger equations and their application to hadronic physics. \textit{Prog.Part.Nucl.Phys.}, \underline{33} (1994) 477-575.
\bibitem{huber} Huber M.Q. Derivation of Dyson-Schwinger equations. physik.uni-graz.at/~mgh/notes/DerivationDSEs.pdf.
\bibitem{robrob} Roberts D.C. Strong QCD and Dyson-Schwinger Equations,   arXiv:1203.5341v1 [nucl-th]
\bibitem{loic} Foissy L.  General Dyson-Schwinger equations and systems,   arXiv:1112.2606v1 [math.RA]
\bibitem{hubermitter} Huber M.Q., Mitter M. CrasyDSE: A framework for solving Dyson-Schwinger equations.
Comput. Phys. Commun. 2012 Nov;183(11):2441-2457.
\bibitem{yeats} Yeats Karen, Rearranging Dyson-Schwinger Equations,  Memoirs of the American Mathematical Society
2011; 82 pp;
\bibitem{davidhugo} Campagnari D., Reinhardt H. Variational and Dyson-Schwinger equations of Hamiltonian quantum chromodynamics
\textit{Phys. Rev. D} \textbf{97}(2018), 054027.
\bibitem{wilrein} Wilson P., Reinhardt H. The Coulomb gauge ghost Dyson-Schwinger equation, \textit{Phys.Rev. D}\textbf{82}:125010, arXiv:1007.2583[hep-ph](2010).
\bibitem{kreimer} Kreimer D. A Lecture series: DYSON-SCHWINGER EQUATIONS, https://www2.mathematik.hu-berlin.de/~kreimer/wp-content/uploads/SkriptDSE.pdf
\bibitem{flyvbjorg} Flyvbjerg H. Dyson-Schwinger equations for the nonlinear sigma model: perturbative solution on a finite lattice., J. of Phys. A: Mathematical and General, 22(1989)3393.
\bibitem{rvsw} Reinhard A., von Smekal L., Watson  P. The Kugo-Ojima Confinement Criterion from Dyson-Schwinger Equations, \textit{Workshop on Dynamical Aspects of the QCD Phase Transition (2001 : Trento, Italy)}, http://arxiv.org/abs/hep-ph/0105142.
\bibitem{bowm} P.O. Bowman, U. M. Heller, D. B. Leinweber, M. B. Parappilly, and A. G. Williams\textit{Phys. Rev.}\textbf{D70} (2004) 034509, hep-lat/0402032.
\bibitem{stenil} A. Sternbeck, E. M. Ilgenfritz, M. Mueller-Preussker, and A. Schiller \textit{Phys. Rev.} \textbf{D72} (2005) 014507, hep-lat/0506007.

\end{thebibliography}
\end{document}